\documentclass[sigconf,nonacm]{acmart}

\AtBeginDocument{%
  }

\copyrightyear{2024}
\acmYear{2024}
\setcopyright{cc}
\acmConference[ArXiv '24]{ArXiv}{2024}{La Jolla, CA, USA}
\acmBooktitle{}
\acmDOI{}
\acmISBN{}

\usepackage{listings}
\usepackage{xcolor}
\usepackage{xspace}
\usepackage[inline,shortlabels]{enumitem}
\usepackage{multirow}
\usepackage{calc}
\usepackage{hyperref}
\usepackage{acmart-taps}

\newcommand{\ie}{\emph{i.e.\@}\xspace}
\newcommand{\eg}{\emph{e.g.\@}\xspace}

\newcommand{\mypara}[1]{\paragraph{#1}}
\def\paragraph#1{\smallskip\noindent\textbf{\textit{#1.}}\xspace}

\newcommand{\scode}[1]{{\texttt{\small #1}}}

\newcommand{\tname}[1]{\textsc{#1}\xspace}
\newcommand{\pb}{Projection Boxes\xspace}
\newcommand{\TOOL}{\tname{Leap}}
\newcommand{\tool}{\TOOL}
\newcommand{\snippy}{\tname{SnipPy}\xspace}
\newcommand{\outputBox}{Output Panel\xspace}
\newcommand{\lp}{Live Programming\xspace}
\newcommand{\panel}{Suggestion Panel\xspace}

\newcommand{\studyone}{Fixed-Prompt\xspace}
\newcommand{\studytwo}{Open-Prompt\xspace}
\newcommand{\pandas}{Pandas\xspace}
\newcommand{\bigram}{Bigram\xspace}
\newcommand{\aoc}{String Rewriting\xspace}
\newcommand{\boxplot}{Box Plot\xspace}
\newcommand{\GroupPB}{LP\xspace}
\newcommand{\GroupNoPB}{No-LP\xspace}

\newcommand{\preview}{\textcolor{blue}{\small \texttt{Preview}}\xspace}

\definecolor{mjcolor}{rgb}{0.1,0.4,0.1}

\newcommand{\medpb}{\mathit{median}_{\mathit{\GroupPB}}}
\newcommand{\mednpb}{\mathit{median}_{\mathit{\GroupNoPB}}}

\begin{document}

\title{Validating AI-Generated Code with Live Programming}

\author{Kasra Ferdowsi}
\authornote{The first two authors, listed alphabetically, contributed equally.}
\email{kferdows@ucsd.edu}
\orcid{0000-0003-3924-8137}
\affiliation{%
  \institution{UC San Diego}
  \city{San Diego}
  \state{CA}
  \country{USA}
  \postcode{92093}
}

\author{Ruanqianqian (Lisa) Huang}
\authornotemark[1]
\email{r6huang@ucsd.edu}
\orcid{0000-0002-4242-419X}
\affiliation{%
  \institution{UC San Diego}
  \city{San Diego}
  \state{CA}
  \country{USA}
  \postcode{92093}
}
\author{Michael B. James}
\email{m3james@ucsd.edu}
\orcid{0009-0007-2219-9355}
\affiliation{%
  \institution{UC San Diego}
  \city{San Diego}
  \state{CA}
  \country{USA}
  \postcode{92093}
}

\author{Nadia Polikarpova}
\email{npolikarpova@ucsd.edu}
\orcid{0000-0001-5571-173X}
\affiliation{%
  \institution{UC San Diego}
  \city{San Diego}
  \state{CA}
  \country{USA}
  \postcode{92093}
}

\author{Sorin Lerner}
\email{lerner@cs.ucsd.edu}
\orcid{0000-0003-3957-0628}
\affiliation{%
  \institution{UC San Diego}
  \city{San Diego}
  \state{CA}
  \country{USA}
  \postcode{92093}
}

\renewcommand{\shortauthors}{Ferdowsi \& Huang et al.}

\begin{abstract}
AI-powered programming assistants are increasingly gaining popularity, with GitHub Copilot alone used by over a million developers worldwide. These tools are far from perfect, however, producing code suggestions that may be incorrect in subtle ways. As a result, developers face a new challenge: \emph{validating} AI's suggestions. This paper explores whether Live Programming (LP), a continuous display of a program's runtime values, can help address this challenge. To answer this question, we built a Python editor that combines an AI-powered programming assistant with an existing LP environment. Using this environment in a between-subjects study ($N=17$), we found that by lowering the cost of validation by execution, LP can mitigate over- and under-reliance on AI-generated programs and reduce the cognitive load of validation for certain types of~tasks.
\end{abstract}





\maketitle

\section{Introduction}\label{sec:intro}

Recent advances in large language models have given rise to AI-powered code suggestion tools
like GitHub Copilot~\cite{copilot}, Amazon CodeWhisperer~\cite{whisperer}, and ChatGPT~\cite{chatgpt}.
These \emph{AI programming assistants} are changing the face of software development,
automating many of the traditional programming tasks,
but at the same time introducing \emph{new tasks} into the developer's workflow---%
such as prompting the assistant and reviewing its suggestions~\cite{barkeGroundedCopilot2022,mozannar2022reading}.
Development environments have some catching up to do
in order to provide adequate tool support for these new tasks.

In this paper, we focus on the task of \emph{validating} AI-generated code,
\ie, deciding whether it matches the programmer's intent.
Recent studies show that validation is a bottleneck for AI-assisted programming:
according to~\citet{mozannar2022reading}, it is the \emph{single most prevalent activity} when using AI code assistants,
and other studies~\cite{Vaithilingam_2022,liang2023understanding,wang2023investigating,birdTakingFlightCopilot2022} 
report programmers having trouble evaluating the correctness of AI-generated code.
Faced with difficulties in validation, programmers tend to either \emph{under-rely} on the assistant---\ie, lose trust in it---%
or to \emph{over-rely}---\ie, blindly accept its suggestions~\cite{Weisz_2021,rossProgrammerAssistantConversational2023,vasconcelosExplanationsCanReduce2023,tang2023empirical};
the former can cause them to abandon the assistant altogether~\cite{barkeGroundedCopilot2022},
while the latter can introduce bugs and security vulnerabilities~\cite{perry2022users}.
These findings motivate the need for better validation support in AI-assisted programming environments.

This paper investigates the use of \emph{Live Programming} (LP)~\cite{hancock2003real,victorLearnableProgramming2012,tanimoto2013perspective}
as a way to support the validation of AI-generated code.
LP environments, such as Projection Boxes~\cite{lerner2020pb},
visualize runtime values of a program in real-time without any extra effort on the part of the programmer.
We hypothesize that these environments are a good fit for
validation,
since LP has been shown to encourage more frequent testing~\cite{cabreraProgramsPalmYour2019}
and facilitate bug finding~\cite{zhaoODENLiveProgramming2022} 
and program comprehension~\cite{delineGlindaSupportingData2021, delineSupportingExploratoryData2015, campusanoDoesLiveProgramming2016}.
On the other hand, validation of AI-generated code is a new and unexplored domain in program comprehension
that comes with its unique challenges,
such as multiple AI suggestions for the programmer to choose from,
and frequent context switches between prompting, validation, and code authoring~\cite{mozannar2022reading}, 
which cause additional cognitive load~\cite{wang2023investigating}.
Hence, the application of LP to the validation setting warrants a separate investigation.

To this end, we constructed a Python environment that combines an existing LP environment~\cite{lerner2020pb}
with an AI assistant similar to Copilot's multi-suggestion pane.
Using this environment, we conducted a between-subjects experiment ($N=17$)
to evaluate how the availability of LP affects users' effectiveness and cognitive load in validating AI suggestions.
Our study shows that \lp facilitates validation through \emph{lowering the cost of inspecting runtime values};
as a result, participants were more successful in evaluating the correctness of AI suggestions
and experienced lower cognitive load in certain types of tasks.

\begin{figure*}[t]
	\centering
	\includegraphics[width=\linewidth]{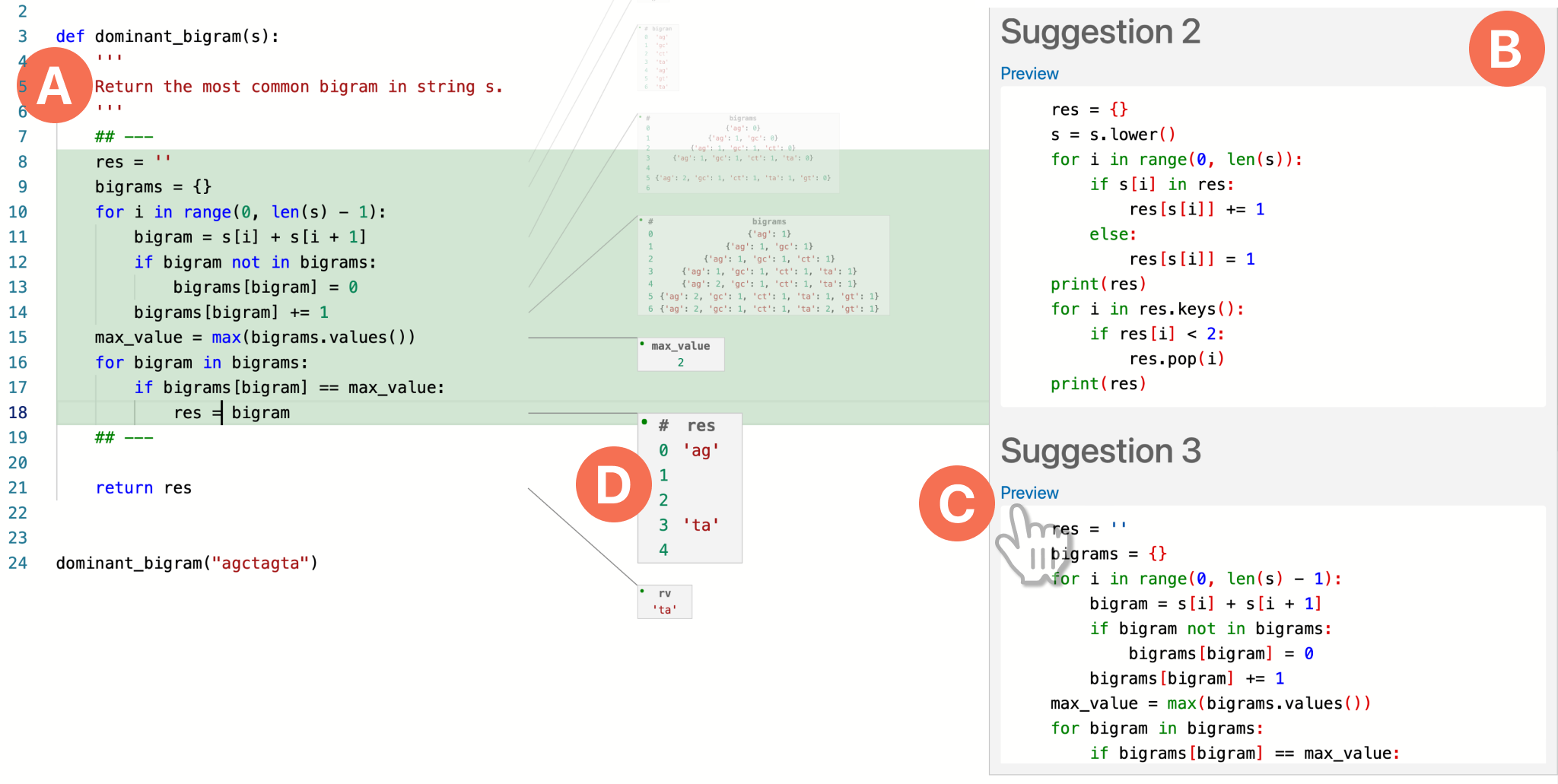}
    \caption[]{
    L\textsc{eap} is a Python environment that enables validating AI-generated code suggestions via \lp.\\
	\raisebox{-2pt}{\includegraphics[width=9.3pt]{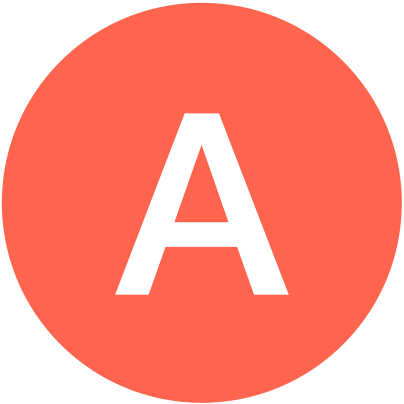}} Users prompt the AI assistant via comments and/or code context.
    \raisebox{-2pt}{\includegraphics[width=9.3pt]{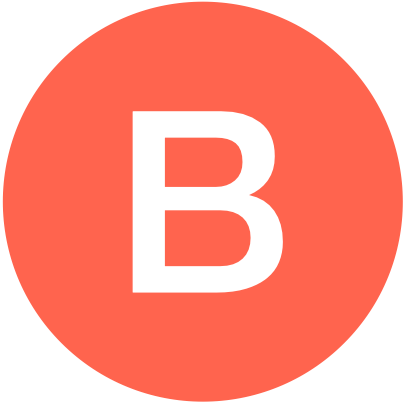}} The \panel shows the AI-generated suggestions.
    \raisebox{-2pt}{\includegraphics[width=9.3pt]{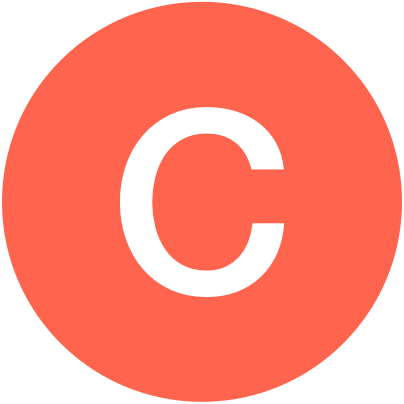}} Pressing a \preview button inserts the suggestion into the editor.
    \raisebox{-2pt}{\includegraphics[width=9.3pt]{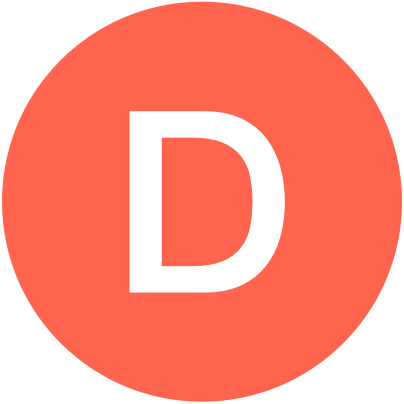}} Users can inspect the runtime behavior of the suggestion in Projection Boxes~\cite{lerner2020pb},
	which are updated continuously as the user edits the code.}
    \Description{
    Screenshot of editor for live evaluation of ai-generated programs.
	There are three visual columns of content.
	The left column shows an in-progress Python function being written, dominant_bigram, and a large A next to the definition and doctstring.
	Inside the function is a highlighted block of code.
	The visual middle column shows Projection Boxes, one for each executed line of code in the dominant_bigram function.
	The emphasized projection box shows runtime values of a variable assignment and a line connecting it to the line of code, with a D label.

	The right-most column shows the Suggestions Panel with two different code suggestions.
	Each code suggestion has a preview button above it.
	There is a B label on the Suggestion Panel.

	A cursor is on the preview button for the second suggestion, which matches the highlighted code in the editor.
	There is a C label on the preview button.
}
    \vspace{-.7em}
	\label{fig:overview}
\end{figure*}

\section{Related Work}\label{sec:related}
\paragraph{Validation of AI-Generated Code} A rapidly growing body of work analyzes how users interact with AI programming assistants.
Studies show that programmers spend a significant proportion of their time validating AI suggestions~\cite{barkeGroundedCopilot2022,mozannar2022reading, birdTakingFlightCopilot2022}.
Moreover, a large-scale survey~\cite{liang2023understanding} indicates that 23\% of their respondents \emph{have trouble evaluating correctness of generated code},
which echoes the findings of lab studies~\cite{Vaithilingam_2022,barkeGroundedCopilot2022} and a need-finding study~\cite{wang2023investigating},
where participants report difficulties understanding AI suggestions and express a desire for better validation support.
%
\citet{barkeGroundedCopilot2022} and \citet{liang2023understanding} find that
programmers use an array of validation strategies,
and the prevalence of each strategy is \emph{closely related to its time cost}.

Specifically, despite the help of execution techniques built into the IDE for validating AI suggestions~\cite{tang2023empirical},
execution is used less often than quick manual inspection or type checking because it is more time-consuming~\cite{barkeGroundedCopilot2022, liang2023understanding}
and interrupts programmers' workflows~\cite{wang2023investigating}.
The lack of validation support designed for AI-assisted programming, as \citet{wang2023investigating} identify,
leads to a higher cognitive load in reviewing suggestions.
The high cost of validating AI suggestions, according to some studies%
~\cite{Weisz_2021,rossProgrammerAssistantConversational2023,vasconcelosExplanationsCanReduce2023},
can lead to both \emph{under-reliance}---lack of trust---and \emph{over-reliance}---uncritically accepting wrong code---%
on the part of the programmer.

Comparatively fewer existing papers explore interface designs to support validation of AI-generated code:
\citet{rossProgrammerAssistantConversational2023} investigates a conversational assistant that allows programmers to ask questions about the code,
while \citet{vasconcelos2022generation} targets over-reliance by highlighting parts of generated code that might need human intervention;
our work is complementary to these efforts in that it focuses on facilitating validation by execution.

\paragraph{Validation in Program Synthesis}
Another line of related work concerns the validation of code generated by \emph{search-based} (non-AI-powered) program synthesizers.
Several synthesizers help users validate generated code by proactively displaying its outputs~\cite{wrex,regae,hplus}
and intermediate trace values~\cite{resl},
although none of them use a 
LP environment.
The only system we are aware of that combines LP and program synthesis is \snippy~\cite{ferdowsi2020snippy},
but it uses LP to help the user specify their intent rather than validate synthesized code.

\paragraph{Live Programming}
%
%
Live Programming (LP) provides immediate feedback on code edits,
often in the form of visualizations of the runtime state~\cite{hancock2003real, victorLearnableProgramming2012, tanimoto2013perspective}.
Some quantitative studies find that programmers with LP find more bugs~\cite{zhaoODENLiveProgramming2022},
fix bugs faster~\cite{kramerHowLiveCoding2014},
and test a program more often~\cite{cabreraProgramsPalmYour2019}.
Others find no effect in knowledge gain~\cite{huangInvestigatingImpactUsing2022}
or efficiency in code understanding~\cite{campusanoDoesLiveProgramming2016}.
Still, qualitative evidence points to the helpfulness of LP for program comprehension~\cite{delineGlindaSupportingData2021, delineSupportingExploratoryData2015, campusanoDoesLiveProgramming2016}
and debugging~\cite{kangOmnicodeNoviceOrientedLive2017, huangInvestigatingImpactUsing2022}.
In contrast to these studies, which evaluate the effectiveness of LP for comprehending and debugging \emph{human-written} code,
our work investigates its effectiveness for validating \emph{AI-generated} code,
a setting that comes with a number of previously unexplored challenges~\cite{mozannar2022reading, wang2023investigating}.

\section{LEAP: The Tool Used in the Study}\label{sec:system}

\begin{sloppypar}
To study how \lp affects the validation of AI-generated code,
we implemented \tool (\textbf{L}ive \textbf{E}xploration of \textbf{A}I-Generated \textbf{P}rograms),
a Python environment that combines an AI assistant with LP.
%
This section demonstrates \tool via a usage example and discusses its implementation.
\end{sloppypar}

\paragraph{Example Usage}
Naomi, a biologist, is analyzing some genome sequencing data using Python.
As part of her analysis, she needs to find the most common bigram (\ie, two-letter sequence) in a DNA strand.%
\footnote{This is one of the programming tasks from our user study,
and each of Naomi's interactions with \TOOL has been observed in some of our participants.}
To this end, she creates a function \scode{dominant\_bigram} (line 3 in \autoref{fig:overview});
she has a general idea of what this function might look like,
but she decides to use \TOOL to help translate her idea into code.

\smallskip
\begin{enumerate}[nosep, leftmargin=13pt]
	\item[]{\hspace*{-13pt}\raisebox{-2pt}{\includegraphics[width=9.3pt]{img/A.png}}} {%
	Naomi adds a docstring (line 5), which conveys her intent in natural language,
	and a test case (line 24), which will help her validate the code.
	With the cursor positioned at line 7, she presses \raisebox{-1.5pt}{\includegraphics[height=10pt]{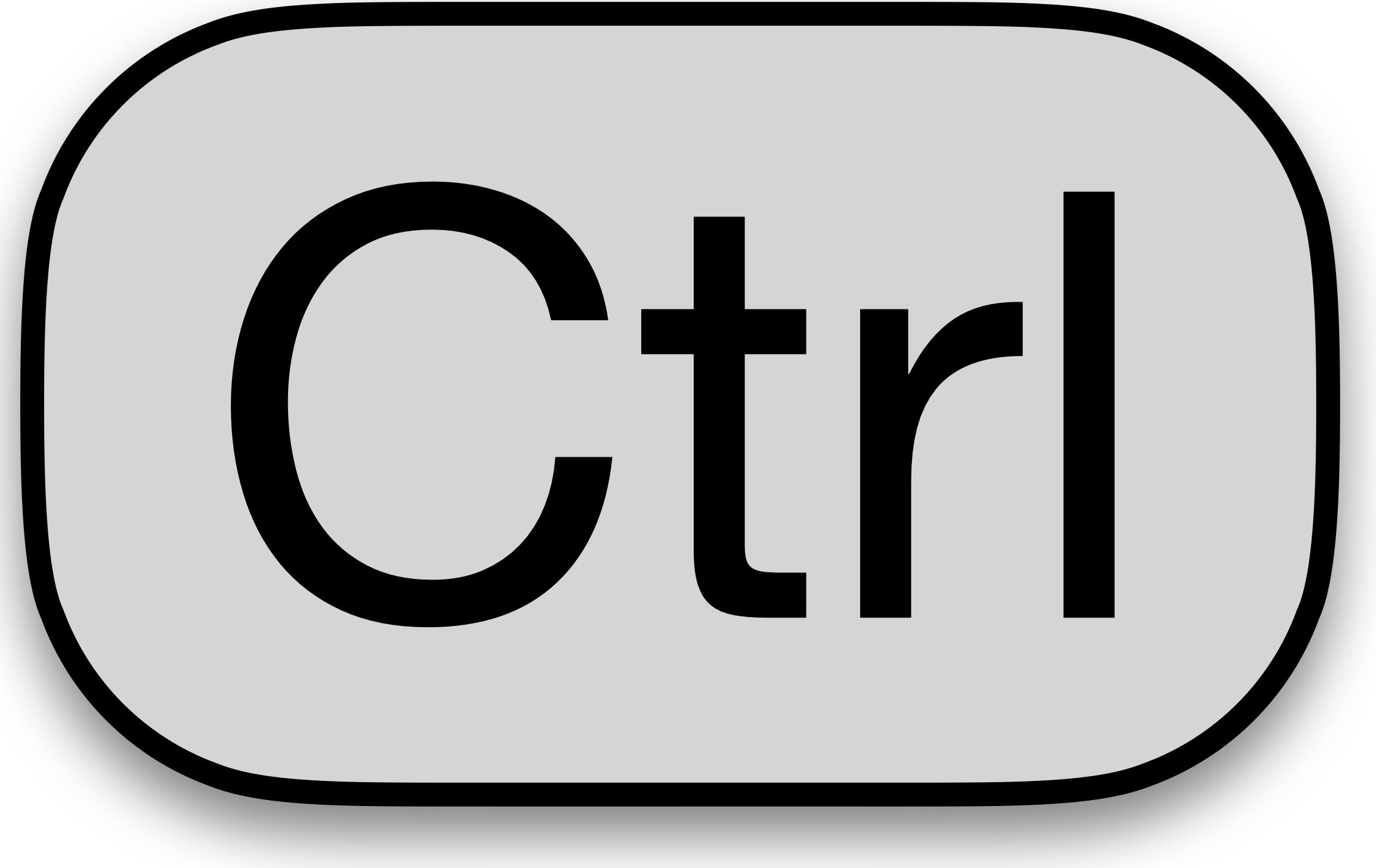}} and \raisebox{-1.5pt}{\includegraphics[height=10pt]{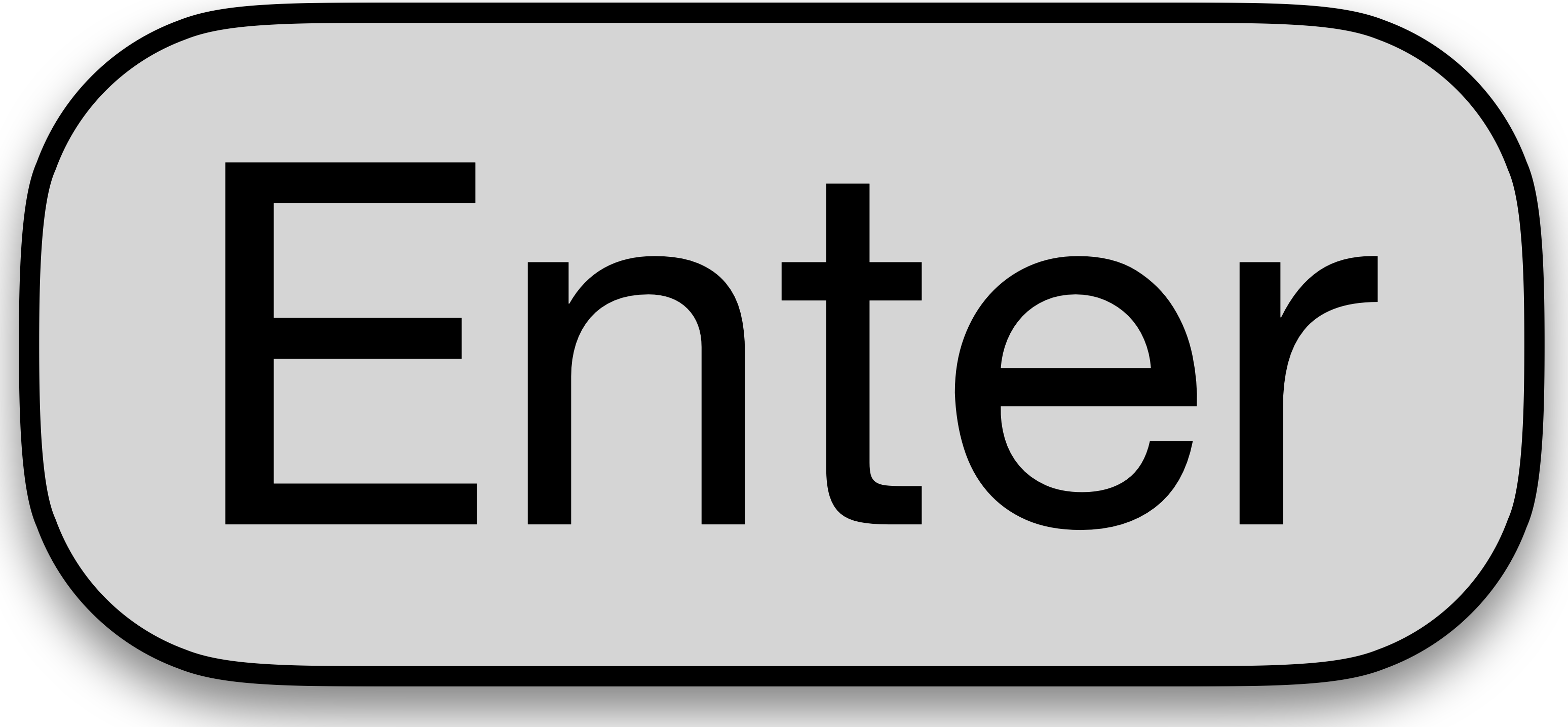}} to ask for suggestions.
	}

	\item[]{\hspace*{-13pt}\raisebox{-2pt}{\includegraphics[width=9.3pt]{img/B.png}}} {%
	Within seconds, a panel opens on the right containing five AI-generated code suggestions;
	Naomi quickly skims through all of them.
	The overall shape of Suggestion 3 looks most similar to what she has in mind:
	it first collects the counts of all bigrams into a dictionary,
	and then iterates through the dictionary to pick a bigram with the maximum count.
	}

	\item[]{\hspace*{-13pt}\raisebox{-2pt}{\includegraphics[width=9.3pt]{img/C.png}}} {%
	Naomi 
	tries
	this suggestion, pressing its \preview button;
	\TOOL inserts the code into the editor and highlights it (lines~8-18).
	}

	\item[]{\hspace*{-13pt}\raisebox{-2pt}{\includegraphics[width=9.3pt]{img/D.png}}} {%
	As soon as the suggestion is inserted, Projection Boxes~\cite{lerner2020pb} appear,
	showing runtime information at each line in the 
	code.
	Inspecting intermediate values helps Naomi understand what the code is doing step by step.
	When she gets to line 18,
	she realizes that the dictionary actually has \emph{two} dominant bigrams with the same count,
	and the code returns \emph{the last one}.
	She realizes this is not what she wants:
	instead, she wants to select the dominant bigram that comes first alphabetically (\scode{ag} in this~case).
	}
\end{enumerate}

\smallskip
\noindent One option Naomi has is to try other suggestions.
She clicks on the \preview button for Suggestion 2; 
\TOOL then inserts Suggestion 2 into the editor, in place of the prior suggestion,
and the \pb update instantly to show its behavior.
Naomi immediately notices that Suggestion 2 throws an exception inside the second loop,
so she abandons it and goes back to Suggestion 3, which got her closer to her goal.

To fix Suggestion 3, Naomi realizes that she can accumulate all dominant bigrams in a list, sort the list, and return the first element.
She does not remember the exact Python syntax for sorting a list,
so she tries different variations---%
including \scode{l = l.sort}, \scode{l = l.sort()}, \scode{l = sort(l)}, \scode{l = l.sorted()}, and so on.
Fortunately, \TOOL's support for LP allows her to get instant feedback on the behavior of each edit,
so she iterates quickly to find 
one correct option: \scode{l = sorted(l)}.
Note that Naomi's workflow for using Suggestion 3---validation, finding bugs, and fixing bugs---relies on full LP support,
and would not work in traditional environments like \emph{computational notebooks},
which provide easy access to the final output of a snippet but not the intermediate values or immediate feedback on edits.

\paragraph{Implementation}
To generate code suggestions, \tool uses the \scode{text-davinci-003} model~\cite{gpt3.5},
the largest publicly available code-generating model at the time of our study.
To support live display of runtime values (\autoref{fig:overview}\raisebox{-2pt}{\includegraphics[width=9.3pt]{img/D.png}}),
we built \tool on top of \pb,
a state-of-the-art LP environment for Python~\cite{lerner2020pb} capable of running in the browser.
The code for \tool can be found at \url{https://bit.ly/leap-code}.
As the control condition for our study,
we also created a version of \tool, where \pb are disabled,
and instead the user can run the code explicitly by clicking a \scode{Run} button
and see the output in a terminal-like \outputBox.

\section{User Study}\label{sec:study}

We conducted a between-subjects study to answer the following research questions:
\begin{enumerate}
\item[\textbf{RQ1)}] How does \lp affect over- and under-reliance in validating AI-generated code?
\item[\textbf{RQ2)}] How does \lp affect validation strategies?
\item[\textbf{RQ3)}] How does \lp affect the cognitive load of validating AI-generated code?
\end{enumerate}

\paragraph{Tasks} Our study incorporates two categories of programming tasks,
\textit{\studyone} and \textit{\studytwo} tasks.

In \textit{\studyone tasks}, 
we provide participants with a \emph{fixed set} of five AI suggestions that are
intended to solve the entire problem.
We curated the suggestions by querying Copilot~\cite{copilot} and \tool
with slight variations of the prompt.
\studyone tasks isolate the effects of \lp on validation behavior
by controlling for the quality of suggestions.
We created two \studyone tasks, each with five suggestions:
 (T1) \textit{\bigram}:
	Find the most frequent bigram in a given string, resolving ties alphabetically (same task in~\autoref{sec:system});
 (T2) \textit{\pandas}:
	Given a \scode{pandas} data frame with data on dogs of three size categories (small, medium, and large),
	compute various statistics, imputing missing values with the mean of the appropriate category.
These tasks represent two distinct styles:
\bigram is a purely algorithmic task, while \pandas focuses on using a complex API.
\pandas has two correct AI suggestions (out of five) while \bigram has none,
a realistic scenario that programmers encounter with imperfect models.

In \textit{\studytwo tasks},
participants are free to invoke the AI assistant however they want.
This task design is less controlled than \studyone, but more realistic, thus increasing ecological validity.
We used two \studytwo tasks:
(T3) \textit{\aoc}:
	parse a set of string transformation rules
	and apply them five times to a string;
(T4) \textit{\boxplot}:
	given a \scode{pandas} data frame containing 10 experiment data records,
	create a \scode{matplotlib} box plot of time values for each group,
	combined with a color-coded scatter plot. 
%
%
Both tasks are more complex than the \studyone tasks,
and could not be solved with a single interaction with the AI assistant.

\paragraph{Participants and Groups}
We recruited 17 participants; 
5 self-identified as women, 10 as men, and 2 chose not to disclose.
6 were undergraduate students, 9 graduate students, and 2 professional engineers.
%
Participants self-reported experience levels with Python and AI assistants:
%
2 participants used Python `occasionally', 8 `regularly', and 7 `almost every day';
7 participants declared they had `never' used AI assistants, and 8 used such tools `occasionally'.
%

There were two experimental groups:
``\GroupPB'' participants used \tool with \pb, as described in \autoref{fig:overview};
``\GroupNoPB'' participants used \tool \emph{without} \pb, instead executing programs in a terminal-like \outputBox.
Participants completed both \studyone tasks
and one \studytwo task.
%
We used block randomization~\cite{efird2011blocked} to assign participants to groups while evenly distributing across task order and selection and balancing experience with Python and AI assistants across groups.
The \GroupPB group had 8 participants, and \GroupNoPB had 9.

\paragraph{Procedure and Data}
We conducted the study over Zoom as each participant used \tool in their web browser.
Each session was recorded and included
two \studyone tasks (10 minutes each),
two post-task surveys,
one \studytwo task (untimed),
one post-study survey,
and a semi-structured interview.
A replication package\footnote{\url{https://bit.ly/leap-study-materials}} shows the details of our procedure, tasks, and data collection.%

For \emph{quantitative} analysis,
we performed closed-coding on video recordings of study sessions to determine each participant's \emph{subjective} assessment of their success on the task;
we matched this data against the \emph{objective} correctness of their final code 
to establish whether they succeeded in accurately validating AI suggestions. %
We also measured task duration---proportion of time \panel (\autoref{fig:overview}\raisebox{-2pt}{\includegraphics[width=9.3pt]{img/B.png}}) was in focus---and participants' cognitive load (via five NASA Task Load Index (TLX) questions~\cite{hart1988development}).
We used Mann-Whitney U tests to assess all differences
except for validation success, which we analyzed via Fisher's exact tests.

In addition, we collected \emph{qualitative} data from both \studyone and \studytwo tasks.
%
We noted validation-related behavior and quotes, which we discussed in memoing meetings~\cite{charmaz2014constructing} after the study.
Through reflexive interpretation, we used category analysis~\cite{yanow2017qualitative} to assemble the qualitative data into groups.
We then revisited notes and recordings to iteratively construct high-level categories.

\section{Results}\label{sec:results}

\subsection{RQ1: Over- And Under-Reliance on AI}\label{subsec:results-validation-accuracy}

To investigate if \lp affects over- and under-reliance,
we measured whether participants successfully validated the AI suggestions in the \studyone tasks,
as described below.
We also compared task completion times
and participants' confidence in their solutions (collected through post-task surveys).
However, neither result was significantly different between the two groups,
so we do not discuss them below.%
\footnote{
 	In median times, the \GroupPB group completed the \pandas task faster by 35 seconds ($p=.664, U=31$).
 	For \bigram, \GroupPB participants were slower by 3 minutes and 51 seconds ($p=.583, U=42$),
 	though this difference changes to \emph{faster} by 10 seconds
 	if we exclude those who solved the task incorrectly.
 	For \pandas, both groups had the median ratings of confidence in correctness as ``Agree'' on seen inputs ($p=.784, U=30$) and ``Neutral'' on unseen inputs ($p=.795, U=33$).
 	For \bigram, the \GroupPB group had the median rating of confidence in correctness on seen inputs as ``Agree'', while the \GroupNoPB group had ``Strongly Agree'' ($p=.097, U=19.5$).
 	As for confidence in correctness on unseen inputs, the median for the \GroupPB group was ``Neutral'', and that for the \GroupNoPB group was ``Agree'' ($p=.201, U=22.5$).}

\begin{figure}[t]
	\centering
	\includegraphics[width=\linewidth]{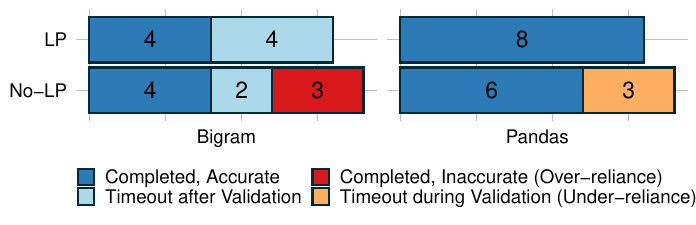}
	\caption{Success in validating AI suggestions across groups for \studyone tasks.
	         ``Completed'' means the participant submitted a solution they were satisfied with by the time limit, and ``Timeout'' means they did not.
			 We deem the validation \emph{successful} if a participant submitted a correct solution (dark blue)
			 or timed out when attempting to fix the correctly identified 
				bugs in their chosen suggestion
			 (light blue).
			}
	\vspace{-.3em}
	\Description {
	Four stacked bar charts arranged in a two by two table.
	The X axis marks Bigram and Pandas.
	The Y axis marks LP and No-LP.
	The possible labels are "Correct, Accurate"; "Timeout after Validation"; "Completed, Inaccurate (Over-reliance)"; and, "Timeout during Validation (Under-reliance)".
	LP and Bigram shows four "Completed, Accurate" and four "Timeout after Validation".
	No-LP and Bigram shows four "Completed, Accurate", two "Timeout after Validation", and three "Completed, Inaccurate (Over-reliance)".
	LP and Pandas shows all eight "Completed, Accurate".
	No-LP and Pandas shows six "Completed, Accurate", and three "Timeout during validation (Under-reliance)".
	}
	\label{fig:consistency}
\end{figure}

\mypara{We found six instances of unsuccessful validation, all from the \GroupNoPB group}
As described in \autoref{sec:study}, we compared subjective and objective assessments of code correctness on the two \studyone tasks,
which resulted in four outcomes:
(1) \emph{Complete and Accurate}, where the participant submitted a correct solution within the task time limit,
(2) \emph{Complete and Inaccurate}, where the participant submitted an incorrect solution
without recognizing the error,
(3) \emph{Timeout after Validation}, where the participant formed an accurate understanding
of the correctness of the suggestions
but reached the time limit before fixing the error in their chosen suggestion,
and (4) \emph{Timeout during Validation}, where the participant reached the time limit
before they had finished validating the suggestions.
We consider (1) and (3) to be instances of \emph{successful validation},
(2) to be an instance of \emph{over-reliance} on the AI suggestions,
and (3) to be an instance of \emph{under-reliance}, as the participant did not
successfully validate the suggestions in the given time.
As \autoref{fig:consistency} shows, we found three instances of over-reliance in the \bigram task
and three instances of under-reliance in the \pandas task,
\emph{all from the \GroupNoPB group},
though the overall between-group difference was not significant
($p=.206$ for both tasks).

\mypara{Participants with over-reliance did not inspect enough runtime behavior}
The three \GroupNoPB participants with over-reliance in \bigram (P5, P12, P15) made a similar mistake:
they accepted one of the mostly-correct suggestions (similar to Suggestion 3 in \autoref{sec:system})
and failed to notice that ties were not resolved alphabetically.
%
Among the three participants, P5 did not run their code at all.
P12 and P15 both tested \emph{only one} suggestion on the given input 
and failed to notice the presence of two bigrams of the same count
(and the fact that other suggestions returned different results).
In addition, P15 cited \emph{``reading the comments on what it was doing''}
as a key factor for choosing the suggestion they did.
That suggestion began with a comment 
stating that it resolved ties alphabetically, but the following code did not do so.

\mypara{Participants with under-reliance lacked affordances for inspecting runtime behavior}
The three \GroupNoPB participants who under-relied on AI suggestions (P7, P9, P15) 
tried to use runtime values for validation but struggled with doing so.
P9 previewed and ran multiple suggestions
but did not add any \scode{print} statements to the code, 
and so they could only see the output of one of the suggestions, 
which ended in a \scode{print} statement.
P15 ran all suggestions
and did add a \scode{print} statement to each to inspect the final return value,
but the need to change the \scode{print} statement and re-run each time made this process difficult,
and they lost track of which suggestions they considered the most promising, 
saying \emph{``I forgot which ones looked decent.''}
Finally, P7's strategy was to print the output of subexpressions from various suggestions
in order to understand their behavior and combine them into a single solution,
but this was time-consuming, so they did not finish.

\subsection{RQ2: Validation Strategies}\label{sec:results:strategy}

Our participants had access to two validation strategies:
\emph{examination} (reading the code) and \emph{execution} (inspecting runtime values).
The general pattern we observed was that participants first did some amount of examination inside the \panel---%
ranging from a quick glance to thorough reading---%
and then proceeded to preview zero or more suggestions,
performing further validation by execution inside the editor.
To this end, \GroupNoPB participants in most tasks ran the code
and added \scode{print} statements for both final and intermediate values;
\GroupPB participants in all tasks inspected both final and intermediate runtime values in \pb
(by moving the cursor from line to line to bring different boxes into focus),
and occasionally added \scode{print} statements to see variables not shown by default.
Below we discuss notable examples of validation behavior,
as well as differences between the two groups and across tasks.

\begin{figure}
	\centering
	\includegraphics[width=\linewidth]{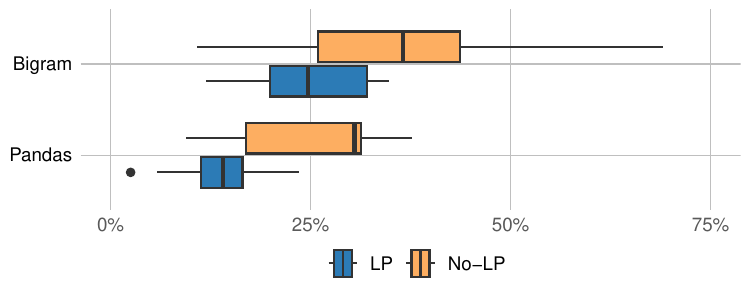}
	\caption{Percentage of time spent in the \panel across the two groups for \studyone tasks.}
	\vspace{-1em}
	\Description {
	Four horizontal box plots, grouped into two of two.
Along the Y axis, top down it is Bigrams then Pandas.
The X axis ranges from 0
The boxplots are ordered No-LP then LP.

The data is as follows:
Task Condition Min Q1 Med Q3 Max
Bigram No-LP 10.8 25.9 36.6 43.6 69.1
Bigram LP 12.0 19.9 24.7 32.0 34.8
Pandas No-LP 9.47 17.0 30.5 31.3 37.7
Pandas LP 2.51 11.3 14.1 16.5 23.5
	}
	\label{fig:focus}
\end{figure}

\begin{sloppypar}
\paragraph{\GroupPB participants spent less time reading the code}
We use the time the \panel was in focus as a proxy for examination time;
\autoref{fig:focus} shows this time as a percentage of the total task duration.
The \GroupNoPB group spent more time in the \panel compared to \GroupPB for both \studyone tasks.
The difference is significant in the \pandas task ($p=.02, U=11, \medpb=14.05\%, \mednpb=30.47\%$)
but not in \bigram ($p=.14, U=20, \medpb=24.70\%, \mednpb=36.57\%$).
We also collected this data for the \studytwo tasks,
although it should be interpreted with caution
due to the unstructured nature of the tasks
(\eg, participant engagement with the assistant and suggestion quality varied).
The results are consistent with the \studyone tasks---\ie, \GroupNoPB participants spent more time in the \panel---%
but the difference is not significant, and the effect in \boxplot is very small
($p=.14, U=3.5, \medpb=6.25\%, \mednpb=15.49\%$ for \aoc;
$p=.67, U=6, \medpb=8.10\%, \mednpb=8.70\%$ for \boxplot).	
\end{sloppypar}

\paragraph{Participants relied on runtime values more in API-heavy, one-off tasks}
According to \autoref{fig:focus}, both groups 
spent more time examining the code in \bigram, 
while in \pandas they jumped to execution more immediately
($\mathit{median}_{\mathsf{\pandas}}=16.96\%, \mathit{median}_{\mathsf{\bigram}}=31.67\%, p=.04, U=206$).
This difference in validation strategies between the two tasks 
was also reflected in the interviews.
For example, P1 described their strategy for \pandas as follows:
\emph{``I didn't look too closely in the actual code, I was just looking at the runtime values on the side.''}
%
Instead, in \bigram, participants cared more about the code itself,
preferring suggestions based on their expected algorithm, data structure, or style
(\eg P15 \emph{``was really looking for the dictionary aspect''}),
with the most popular attribute being ``short''/``readable'',
cited by 10 out of 17 participants.
One explanation participants gave for the difference in behavior is that \pandas is an API-heavy task,
and when dealing with unfamiliar APIs, reading the code is just not very helpful:
\emph{``When it's using more jargony stuff that doesn't translate directly into words in your brain,
then seeing the preview makes it clearer''} (P3).
Another explanation they gave
is that \pandas was perceived by the participants as a \emph{one-off} task,
\ie, it only needed to work on the one specified input,
whereas \bigram was perceived as \emph{general},
\ie, it needed to work on \emph{``any sort of string~[\ldots]~not only~[\ldots]~the specific string that was tested''} (P3);
this was not explicit in the instructions,
but in retrospect it is a reasonable assumption, given the problem domains and structure of the starter code.
On the other hand, some \GroupPB participants conjectured that with more familiarity with \lp,
they would rely on runtime values more, even in tasks like \bigram:
\emph{``If I were to use this tool again I would preview more immediately,
just because I think I was very focused on whether it produced how I would solve the problem
vs. whether it solved the problem correctly''} (P4).

\paragraph{\GroupPB participants benefitted from visualizing intermediate values}
We looked into the validation strategies used in \bigram 
to identify the tie-resolution issue in AI suggestions
(excluding P17 because they wrote the code from scratch).
In the input we provided,
it was hard to identify the most common bigram at a glance,
which made it difficult to validate suggestions just by looking at the final result.
\emph{Five out of eight} \GroupPB participants 
found the issue by inspecting \emph{intermediate values}
and noticing that multiple bigrams in the input have the same count
(the other three relied on custom test cases and code examination).
In the \GroupNoPB group,
three out of eight participants failed to identify the issue
and of the remaining five who succeeded, \emph{only one} (P6) relied on intermediate values to do so.
%
In addition, multiple \GroupPB participants (P1, P3, P4) mentioned the usefulness of intermediate values in the interview,
especially for long suggestions.
P1 said:
	\emph{
	``Because it's a block of text as a suggestion, having projection boxes is more important [\ldots]
	my thought was `let me go line by line to see what is going on'.''
	}
%
In contrast, a \GroupNoPB participant (P9) remarked that they \emph{``had to really look through the code and try to visualize it in [their] mind.''}

\paragraph{\GroupPB participants used liveness features for validation and debugging}
For validation, \GroupPB participants made use of full liveness,
\ie, the ability to see the immediate effects of their edits.
\emph{Five} participants in \pandas added auxiliary calculations 
to double-check the correctness of the final output,
\eg, the mean of specific cells in the input table, comparing it to the output table.
%
When it comes to debugging, \GroupPB participants made multiple rounds of trial and error
guided by liveness.
In fact, the example in \autoref{sec:system} was inspired by P4's debugging process in the Bigram task.
Also, in \boxplot, P1 made many repeated edits in an AI suggestion to tune the placement of a label,
guided by error messages and incorrect outputs to figure out the precise usage of an unfamiliar API call.
In the interview, they noted: \emph{``I was definitely using the projections [...] as I was editing the suggestions to see if my intended changes actually were followed through.''}

\begin{figure*}
		\centering
		\includegraphics[width=\linewidth]{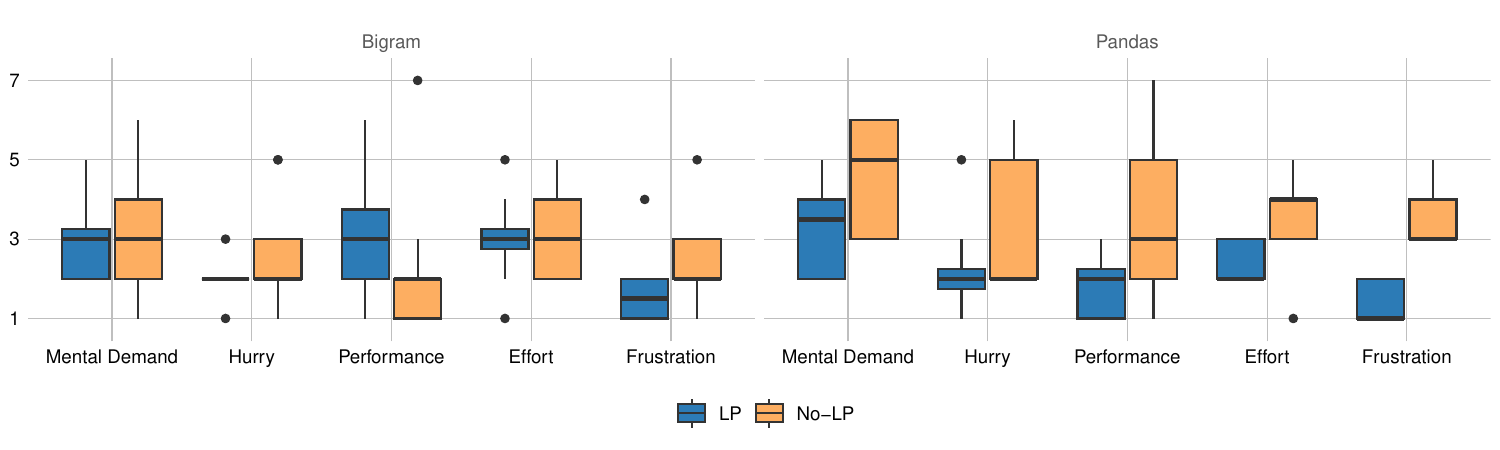}
		\caption{NASA Task Load Index (TLX) results for the \studyone tasks: \bigram on the left, and \pandas on the right. Higher scores indicate higher cognitive load (in case of Performance this means higher failure rate).}
		\vspace{-.3em}
		\label{fig:tlx}
		\Description{
Box Plot of 5 metrics, where each metric has two box plots, one for PB and the other for No-PB.
The Y axis measures from 1 to 7.
The X axis labels from left-to-right are: Mental Demand, Hurry, Performance, Effort, and Frustration.

There are P-values and U-statistics for statistical significance over Mental Demand, Performance, Effort, and Frustration.

Over Mental Demand p-value is .039, u-statistic is 14.5.
Over Performance p-value is .048, u-statistic is 15.5.
Over Effort p-value is .015, u-statistic is 11.
Over Frustration p-value is .0004 u-statistic is 0.

Raw data table follows:
Condition Question        Min    Q1   Med    Q3   Max
Pandas PB        Mental Demand     2  2      3.5  4        5
Pandas PB        Hurry             1  1.75   2    2.25     5
Pandas PB        Performance       1  1      2    2.25     3
Pandas PB        Effort            2  2      2    3        3
Pandas PB        Frustration       1  1      1    2        2
Pandas No-PB     Mental Demand     3  3      5    6        6
Pandas No-PB     Hurry             2  2      2    5        6
Pandas No-PB     Performance       1  2      3    5        7
Pandas No-PB     Effort            1  3      4    4        5
Pandas No-PB     Frustration       3  3      3    4        5
		}
\end{figure*}

\subsection{RQ3: Cognitive Load in Validation}\label{subsec:tlx}

\paragraph{\GroupPB participants experienced significantly lower cognitive load in the \pandas task but not the \bigram task}
In \pandas, 
\GroupPB participants experienced significantly lower cognitive load in four out of five aspects of NASA-TLX~\cite{hart1988development}: mental demand ($p=.039, U=14.5$), performance ($p=.048, U=15.5$), effort ($p=.015, U=11$), and frustration ($p=.0004, U=0$).
We find no significant differences in responses to \bigram, 
but \GroupPB participants reported slightly higher \emph{performance} measures ($\medpb=3, \mednpb=2$), 
which stand for higher failure rates.

\paragraph{Participants found LP helpful in distinguishing between multiple suggestions}
Participants from both the \GroupNoPB (P9, P14, P17) and \GroupPB (P3, P16) groups commented on the utility of seeing multiple suggestions at once:
\emph{``[Seeing multiple suggestions] gave me different ways to look at the code and gave me different ideas''} (P9) and \emph{``multiple suggestions gave points of comparison that were useful''} (P14).
However, some \GroupNoPB participants (P6, P7, P15, P17) said they found the suggestions hard to distinguish.
They noted the difficulty of differentiating just by reading the code because \emph{``the suggestions [were] all almost the same thing''} (P7), and observed that \emph{``the tool did not really help with choosing between suggestions''} (P15).
In comparison, some in the \GroupPB group (P1, P16) 
specifically commented that \lp was helpful in distinguishing and choosing between multiple code suggestions; P1 said:
	\emph{
	``Being able to preview, edit, and look at the projection boxes before accepting a snippet was very helpful when choosing between multiple suggestions.''
	}
As far as we are aware, this is a new application of \lp, specific to AI programming assistants and not previously explored in \lp literature.

\section{Discussion}\label{sec:discussion}

\paragraph{\lp lowers the cost of validation by execution}
Although both \GroupPB and \GroupNoPB participants had access to runtime values as a validation mechanism,
those without LP needed to
examine the code to decide which values to print,
add the \scode{print} statements,
run the code, 
and match each line in the output to the corresponding line in the code.
If they wanted to inspect a different suggestion, they had to repeat this process from the start.
Meanwhile, \GroupPB participants could simply 
preview a suggestion
and get immediate access to all the relevant runtime information,
easily switching between suggestions as necessary.
In other words, LP lowers \emph{the cost}---in terms of both time and mental effort---of access to runtime values.
As a result, we saw \GroupPB participants relied on runtime values more for validation,
as they spent less time examining the code in general---and significantly so for the \pandas task---%
and more often used intermediate values to find bugs in \bigram (\autoref {sec:results:strategy}).
Our findings are consistent with prior work~\cite{barkeGroundedCopilot2022,liang2023understanding},
which demonstrated that programmers more often use validation strategies with lower time costs.
Hence, \emph{by lowering the cost of access to runtime values, \lp promotes validation by execution.}   

\paragraph{The lower cost of validation by execution prevents over- and under-reliance}
As discussed in \autoref{subsec:results-validation-accuracy}, we found six instances of unsuccessful validation in our study,
\emph{all from the \GroupNoPB group}, over-relying on AI suggestions in the \bigram task, and under-relying in \pandas.
We attribute these failures to the high cost of validation by execution:
those who over-relied 
did not inspect the runtime behavior of the suggestions in enough detail,
while those with under-reliance lacked the affordances to do so effectively,
and so ran out of time before they could validate the suggestions.
Our results echo prior findings~\cite{vasconcelosExplanationsCanReduce2023} 
that relate the cost of a validation strategy to its effectiveness in reducing over-reliance on AI.
Prior work has also shown~\cite{ferdowsi2020snippy,wang2023investigating} that programmers often struggle to form an appropriate level of trust in code synthesizers,
	whether AI-based or not;
	our results suggest an important new role for \lp in addressing this challenge.
We conclude that \emph{the lower cost of validation by execution in \lp 
leads to more accurate judgments of the correctness of AI-generated code.}

\paragraph{Validation strategies depend on the task}
\autoref{sec:results:strategy} shows that
participants overall spent significantly more time examining the code in \bigram than in \pandas
and also paid more attention to code attributes in the former.
Participants explained the difference in their validation strategies by two factors:
(1) \pandas contained unfamiliar API calls, the meaning of which they could not infer from the code alone;
and (2) they perceived \pandas as a one-off task, which only had to work on the given input.
We conjecture that (1) is partly due to our participants being LP novices:
as they get more used to the environment,
they are likely to rely on previews more, even if they are not forced into it by an unfamiliar API 
(as P4 mentioned in~\autoref{sec:results:strategy}).
(2), though, is more fundamental:
when dealing with a general task, 
correctness is not all that matters; 
code quality becomes important as well, and LP does not help with that.

In \studytwo tasks,
code examination was less prevalent in the overall task duration,
because in these tasks participants spent a significant amount of time on activities besides validation
(\eg, decomposing the problem and crafting prompts).
It might seem surprising, however, that we did not see any difference in examination time between the two groups in \boxplot,
which is an API-heavy, one-off task, similar to \pandas.
This might be because, in \boxplot, the cost of validation by execution was already low for \GroupNoPB participants:
this task did not require inspecting intermediate values,
because the effects of each line of code were reflected on the final plot in a compositional manner
(\ie, it was easy to tell what each line of code was doing just by looking at the final plot).

In conclusion, \emph{\lp does not completely eliminate the need for code examination
but reduces it in tasks amenable to validation by execution.}

\paragraph{\lp lowers the cognitive load of validation by execution}
In \pandas, 
\GroupPB participants experienced lower cognitive load in four out of five TLX categories (\autoref{subsec:tlx}).
This confirms our hypotheses that LP lowers the cost of validation by execution,
and that \pandas is a task amenable to such validation.
More specifically, we conjecture that, by automating away the process of writing \scode{print} statements, 
LP reduces workflow interruptions,
which were identified as one of the sources of increased cognitive load in reviewing AI-generated code~\cite{wang2023investigating}.

In \bigram,
however,
we did not observe a similar reduction in cognitive load;
in fact, \GroupPB participants reported \emph{higher} cognitive load in the ``performance'' category
(\ie, they perceived themselves as less successful).
Our interpretation is that 
the cognitive load in this task was dominated by debugging and not validation,
and whereas all participants in the \GroupPB group engaged in debugging,
only two-thirds of the \GroupNoPB group did so.
Finally, the higher ``performance'' ratings from the \GroupPB group
were from those who ran out of time trying to fix the code,
and hence were aware that they had failed.
These findings show that \lp by itself does not necessarily help with debugging a faulty suggestion.
As we saw in \autoref{sec:results:strategy},
it can be helpful when the user has a set of potential fixes in mind,
which they can quickly try out and get immediate feedback on.
But when the user does not have potential fixes in mind,
they need to rely on other tools, such as searching the web or using chat-based AI assistants.

From these findings, we conclude that \emph{\lp lowers the cognitive load of validating AI suggestions
when the task is amenable to validation by execution.}

\section{Conclusion and Future Work}\label{sec:conclusion}

We investigated an application of \lp in the domain of AI-assisted programming,
finding that LP can reduce over- and under-reliance on AI-generated code
by lowering the cost of validation by execution. 
Our work highlights new benefits of LP specific to AI-assisted programming, 
such as building appropriate trust in the assistant and helping to choose between multiple suggestions.
Our study is necessarily limited in scope:
we focused on self-contained tasks due to LP's limited support for complex programs~\cite{tanimoto2013perspective, lerner2020pb}
and its need for small demonstrative inputs~\cite{soares2013live}.
We hope that our findings inform future studies on code validation
and motivate further research into AI-LP integration.
To that end, we highlight key opportunities below.

To offer liveness, LP places several burdens on the user.
The user must
provide a complete executable program and
a set of test cases, and then
look through potentially large runtime traces for the relevant information.  
AI may alleviate these burdens by
filling in missing runtime values~\cite{lexecutor} for incomplete programs,
generating test cases~\cite{ticoder,WilsonThomas2023}, 
and predicting the most relevant information to be displayed at each program point.
Looking beyond the validation of newly generated code,
there are also opportunities for AI-LP integration for debugging and code repair~\cite{WilsonThomas2023}.
In combination, AI-LP would tighten the feedback loop of querying and repairing AI-generated code:
users could validate code via LP, 
request repair using the runtime information from LP~\cite{ferdowsi2020snippy},
and further validate the repair in LP.

\begin{acks}
We would like to thank Elena Glassman for
her invaluable feedback on a draft of this work.
This work was supported in part by NSF grant 2107397.
\end{acks}

\balance 
\bibliographystyle{ACM-Reference-Format}
\bibliography{citations}

\end{document}